# Realization of robust boundary modes and non-contractible loop states in photonic Kagome lattices

Jina Ma[1], Jun-Won Rhim[2,3], Liqin Tang[1,4†], Shiqi Xia[1], Haiping Wang[1], Xiuyan Zheng[1], Shiqiang Xia[1,5], Daohong Song[1,4], Yi Hu[1,4], Yigang Li[1], Bohm-Jung Yang[2,3,6], Daniel Leykam[7], and Zhigang Chen[1,4,8*]

[1]*The MOE Key Laboratory of Weak-Light Nonlinear Photonics, TEDA Applied Physics Institute and School of Physics, Nankai University, Tianjin 300457, China*
[2]*Department of Physics and Astronomy, Seoul National University, Seoul 08826, Korea*
[3]*Center for Correlated Electron Systems, Institute for Basic Science (IBS), Seoul 08826, Korea*
[4]*Collaborative Innovation Center of Extreme Optics, Shanxi University, Taiyuan, Shanxi 030006, People's Republic of China*
[5]*Engineering Lab for Optoelectronic Technology and Adv. Manufacturing, Henan Normal University, Xinxiang, 453007, China*
[6]*Center for Theoretical Physics (CTP), Seoul National University, Seoul 08826, Korea*
[7]*Center for Theoretical Physics of Complex Systems, Institute for Basic Science, Daejeon 34126, Republic of Korea*
[8]*Department of Physics and Astronomy, San Francisco State University, San Francisco, California 94132, USA*

†tanya@nankai.edu.cn,  *zgchen@nankai.edu.cn

**Abstract:** Corbino-geometry has well-known applications in physics, as in the design of graphene heterostructures for detecting fractional quantum Hall states or superconducting waveguides for illustrating circuit quantum electrodynamics. Here, we propose and demonstrate a photonic Kagome lattice in the Corbino-geometry that leads to direct observation of non-contractible loop states protected by real-space topology. Such states represent the "missing" flatband eigenmodes, manifested as one-dimensional loops winding around a torus, or lines infinitely extending to the entire flatband lattice. In finite (truncated) Kagome lattices, however, line states cannot preserve as they are no longer the eigenmodes, in sharp contrast to the case of Lieb lattices. Using a continuous-wave laser writing technique, we experimentally establish finite Kagome lattices with desired cutting edges, as well as in the Corbino-geometry to eliminate edge effects. We thereby observe, for the first time to our knowledge, the robust boundary modes exhibiting self-healing properties, and the localized modes along toroidal direction as a direct manifestation of the non-contractible loop states.

Flat band (FB) systems have attracted considerable interest in many different branches of physics in the past decade, providing a flexible platform for studying fundamental phenomena associated with completely dispersionless bands within the whole Brillouin zone [1-11]. Engineered FB structures have now been realized in a variety of physical systems [12], ranging from photonic waveguide arrays [13-22] to synthetic atomic lattices [23, 24], and from metamaterials [25] to cavity polaritons [26], enabling exploration of fundamental effects such as Anderson localization, fractional quantum Hall states, superconductivity and superfluidity, and gap solitons. Recently, flat bands have also been experimentally accomplished in realistic materials forming electronic Lieb and Kagome lattices [27, 28].

The Kagome lattice, essentially a two-dimensional (2D) counterpart of the "pyrochlore" structure [Fig. 1(a)], has served as a prototypical FB model. This lattice has a dispersionless FB touching a dispersive band in momentum space [Fig. 1(b)], and thus can support *compact localized states* (CLSs) when hopping is limited to nearest neighbors [12]. The CLSs, as demonstrated in several experiments with photonic Lieb lattices [14, 15, 18], are linearly dependent so they do not form a complete basis for the FB [22]. The missing states are related to the band touching at the conical intersection, which, as argued by Bergman et al. [29], is protected by real-space topology in a sense that the touching is protected and can only be removed by perturbations that also destroy the FB. The missing states can be manifested as eigenstates whose support is extended along non-contractible loops winding around the entire (toroidal) lattice structure with periodic boundary conditions – the so-called *non-contractible loop states* (NLSs). Although the Kagome lattice has been the focus of numerous studies [30-36], to our best knowledge, the NLSs originally proposed for the Kagome lattice have never been realized, as it is not feasible to have an infinite lattice in experiment. Recently, *line states* (LSs) accounting for the missing states from the FB eigenstates have been observed in a finite Lieb lattice with appropriate open boundaries [22]. These line states, independent from linear superpositions of conventional bulk CLSs [37], can be considered as an indirect illustration of the NLSs. Naively, one would think that such line states, localized in one direction but infinitely extended in the other [Fig. 1(c)], should also exist in a finite-sized Kagome lattice. However, as we shall show in this work, these unconventional FB states can manifest dramatical differences in the FB lattices of different topological structures, leading to a new understanding of the fundamental origin of the NLSs as unique topological entities [38].

In this paper, we demonstrate that a straight-line state cannot be the FB eigenmode in finite Kagome lattices regardless of the lattice edges, in contrast to the case of Lieb lattices. As such, we propose and realize alternative ways to observe the NLSs. Firstly, we demonstrate that the Kagome lattices with desired "cutting" edges, written in a nonlinear crystal with a continuous-wave (cw) laser, can support *robust boundary modes* (RBMs) arising from the NLSs due to bulk-edge correspondence [38]. We observe that such RBMs can preserve during propagation and possess "self-healing" features against perturbation in either initial phase or amplitude, as corroborated by numerical simulations. Secondly, we propose and demonstrate a new method for direct observation of the NLS in a Kagome lattice annularly-shaped in a Corbino-geometry. Although a torus geometry with periodicities along two directions [Fig. 1(d)] is difficult to be realized in real systems, one can still preserve the periodicity at least along one direction by making the Corbino-geometry. (Similar geometry has been employed recently for designing graphene heterostructures and superconducting waveguides [39-41]). By applying such a geometry to the Kagome lattice, we observe evidently the "bulk" modes along the toroidal direction as a direct manifestation of the NLSs.

As depicted in Fig. 1(a), the Kagome lattice has three lattice sites (*A*, *B*, and *C*) per unit cell [see black dashed hexagon in Fig. 1(a)], and every lattice site has four nearest neighbors. For our modeling and experiment, the Kagome structure is in the transverse *x*-*y* plane, and it remains invariant along the longitudinal *z*-direction. To excite a particular FB state, a modulated probe beam is sent to propagate though the lattice along the *z*-direction. As well established already in literature [12-19], light propagation in a photonic lattice in the paraxial approximation is described by a Schrödinger equation: $\partial_z \psi(x,y,z) = \left[\frac{i}{2k_0 n_0}\nabla_\perp^2 + ik_0 \Delta n(x,y)\right]\psi(x,y,z)$, where $\psi(x,y,z)$ is the envelope of the electric field $E(x,y,z)$. The optically-induced refractive-index profile $\Delta n(x,y)$ across the lattice acts as an effective potential for the light field. $k_0$ is the free-space wave number, and $n_0$ is the bulk refractive index. Under the tight binding approximation, the continuous Schrodinger equation can be replaced by a discrete one describing the coupling between the individual lattice sites. By Fourier transforming the tight-binding Hamiltonian into $\boldsymbol{k}$-space, one can obtain:

$$H_k = 2t \begin{pmatrix} 0 & \cos k_1 & \cos k_2 \\ \cos k_1 & 0 & \cos k_3 \\ \cos k_2 & \cos k_3 & 0 \end{pmatrix} \quad (1)$$

where $t$ is the hopping amplitude (or coupling constant) for the nearest-neighbor site. The wave vectors are $k_n = \boldsymbol{k} \cdot \boldsymbol{\alpha}_n$, and $\boldsymbol{\alpha}_n$ is illustrated in Fig. 1(a). Then, we can obtain the eigenvalues of $H_k$, as the energy spectrum with three bands:

$$E_1 = -2t$$

$$E_{2,3} = t\left[1 \pm \sqrt{4(\cos^2 k_1 + \cos^2 k_2 + \cos^2 k_3) - 3}\right] . \quad (2)$$

As seen from Eq. (2), the lattice supports three energy bands - two dispersive ($E_{2,3}$) bands and one flat band ($E_1$) in the bottom [Fig. 1(b)]. The two dispersive bands intersect at two inequivalent Dirac points described by a linear dispersion relation similar to that of the honeycomb lattices [42, 43]. The second dispersive band touches the bottom FB at the center of the first Brillouin zone. In FB lattices, one can always find a set of CLSs as degenerate eigenstates, whose energy is the same no matter where they are located in the lattices. A fundamental CLS (ring mode) [19, 44] of the Kagome lattice is illustrated by the red dashed hexagon in Fig. 1(a), where the wave amplitude remains nonzero only at the six lattice sites. Other kinds of CLSs can be constructed by various linear superposition of the fundamental CLS [14, 15, 18, 19]. However, for the Kagome lattice, it was shown that the CLSs cannot span the flat band completely because one can always find a linear combination of CLSs that vanishes [29]. The missing states were found to be the NLSs, which exhibit novel topology on a torus geometry in real space [Fig. 1(d)] although the system is topologically trivial in momentum space. Recently, it has been shown rigorously in theory that the NLSs exist in a flat band because the Bloch wave function of the flat band is discontinuous at the band crossing point in general [38].

Since a torus geometry is hard to be fabricated experimentally, a natural question is whether the line states should exist in a truncated Kagome lattice under appropriate open boundaries (as for the Lieb lattice [22]). Therefore, we shall consider the FB model in photonic Kagome lattices with open boundaries. A torus representing an infinite system can be terminated into a semi-infinite system with four different edges (terminated in *x*-direction including zigzag edge and zigzag-armchair edge, and in *y*-direction including flat edge and armchair edge). The termination of the Kagome lattice causes the non-contractible loop to break into a line. As typical examples, we consider two cases here: the zigzag-armchair edges for the line oriented in *x*-direction [Figs. 1(e1)] and the armchair edges for the line tilted in *y*-direction [Fig.1(e2)], both of which satisfy the destructive interference condition similar to the case of a finite Lieb lattice [22]. Furthermore, by cutting the torus in both toroidal and poloidal directions along the non-contractible loops, it can be unfolded into a 2D plane with localized modes along its boundary as shown in Figs. 1(f1) and 1(f2), which enables the generation of the RBMs due to bulk-boundary correspondence [38]. Importantly, we shall establish a Corbino-geometry of the Kagome lattice to directly observe the NLS in experiment.

We perform a series of experiments starting with cw-laser-writing of finite-sized Kagome lattices with different edges using the technique established already for photonic Lieb lattices [22]. The principle of the technique is essentially site-to-site optical induction in a bulk photorefractive nonlinear crystal (SBN, with dimensions $5\times10\times5mm^3$). Although the physical mechanism for refractive index change is quite different, the writing method used here is similar to the one with femtosecond laser writing waveguides in glass [45, 46]. A cw laser beam ($\lambda$=532nm) is used to illuminate a spatial light modulator (SLM), which can control the amplitude and phase of the writing beam with reconfigurable input positions. After the SLM, a 4F system is used to guarantee a quasi-nondiffracting zone of the writing beam as it propagates through the biased SBN crystal. Because of the noninstantaneous self-focusing photorefractive nonlinearity, all waveguides remain intact during the one-by-one writing process. To visualize the induced Kagome lattices experimentally, we illuminate a weak extraordinarily polarized quasi-plane wave to probe the waveguide array after writing.

At the back facet of the crystal, we see that the otherwise uniform probe beam becomes discretely guided into the lattice sites, indicating the desired Kagome lattice structure has been established [Figs. 2(a, b)].

Then, we examine the line state formation by illuminating the lattice with a line-shaped probe beam with nearly uniform amplitude [Fig. 2(c1)]. The output after exiting the lattice is shown in Figs. 2(c2-c3), for the lattice termination with (left-right) zigzag-armchair edges as illustrated in Fig.1(e1) and experimentally obtained in Fig. 2(a). Apparently, with opposite phase between adjacent sites, the line beam can preserve after 10mm (about 1.5 coupling length) of propagation through the lattice [Fig. 2(c2)]. For comparison, when adjacent sites are in-phase, the line beam exhibits diffraction in orthogonal direction and cannot remain localized [Fig. 2(c3)]. These observations agree well with the numerical simulation results based on the coupled mode theory [Fig. 2(d1) and Fig. 2(d3)]. However, to show the stability of any localized state, we perform further simulations to a distance much longer than what we can achieve in experiment. Surprisingly, simulation results obtained under the tight-binding condition show clearly that, after a longer propagation distance [e.g., 40mm, as shown in Fig. 2(d2)], the line beam cannot preserve its shape as energy leaks to nearby sites close to the two edges. Likewise, for the lattice termination with (top-bottom) armchair edges as illustrated in Fig.1(e2) and obtained in Fig. 2(b), the experimental results are similar [Figs. 2(e1-e3)] – phase sensitive and seemly localized for the out-of-phase case; but simulation to longer distances shows that the line state is not preserved. This is in sharp contrast with the case of Lieb lattices, in which a line state remains localized in finite lattices terminated appropriately [22], making one wonder if the line state is truly an eigenstate of the finite Kagome lattices.

To clarify this issue, let us analyze the line state for the Kagome lattices of Fig. 2(a). As illustrated in Fig. 3(a), the line beam occupies only $B$ and $C$ sites, and it can be described by $|\psi_{LS}\rangle = c_0(|B_1\rangle - |C_1\rangle + |B_2\rangle - |C_2\rangle + |B_3\rangle - |C_3\rangle + |B_4\rangle)$, where $c_0$ is a normalization constant. Applying the tight-binding Hamiltonian to this line state, we have $H_k|\psi_{LS}\rangle = c_0 t(-|B_1\rangle + 2|C_1\rangle - 2|B_2\rangle + 2|C_2\rangle - 2|B_3\rangle + 2|C_3\rangle - |B_4\rangle) \neq |\psi_{LS}\rangle$, where we have considered only non-vanishing terms. From this

simple calculation, one can see that $|\psi_{LS}\rangle$ is not an eigenmode of the terminated Kagome lattice (open boundary condition) because we obtain different coefficients at two ending sites ($B_1$ and $B_4$) compared with the bulk sites ($C_1$, $B_2$, $C_2$, $B_3$, and $C_3$). If the periodic boundary condition is assumed, identifying $B_1$ and $B_4$, $|\psi_{LS}\rangle$ can become an eigenmode. Actually, one can determine whether the line state can be an eigenmode or not from the configuration of the NLS. For the Lieb lattice [22], there is a zero-amplitude site between neighboring sites of the NLS [see Fig. 3(b)]. As such, if we make an open boundary from the torus geometry along a line passing through the sites with the zero amplitude, the line state obtained by cutting the NLS at this boundary becomes an eigenmode of the truncated Lieb lattice, so long the lattice boundary consists of dangling sites (or bearded edges). However, in the case of the Kagome lattice, its NLS has non-vanishing amplitudes at all sites along the non-contractible loop. This means that the NLS of the periodic Kagome lattice satisfies the eigenvalue equation by exchanging its amplitudes at the neighboring sites through the hopping process, unlike the case of the Lieb lattice. Therefore, the line state obtained by cutting the NLS cannot be an eigenstate of the Kagome system with an open boundary as it eliminates one of the hopping processes. To overcome such an open boundary condition issue, one could design a decorated boundary in the Kagome lattice. For instance, by attaching two additional dangling sites at the boundary [denoted by $L_{1,2}$ and $R_{1,2}$ in Fig. 3(c)], the line state can become an eigenstate provided that the hopping parameter for the "dashed bond" is 3 times larger but of opposite sign with respect to that of the "solid bond" at the boundary. This can be readily shown by applying the nearest-neighbor Hamiltonian to the initial state in Fig. 3(c), which now is described by $|\psi_0\rangle = c_0(|L_1\rangle + |L_2\rangle - |B_1\rangle + |C_1\rangle - |B_2\rangle + |C_2\rangle - |B_3\rangle + |C_3\rangle - |B_4\rangle + |R_1\rangle + |R_2\rangle)$ with the hopping parameters chosen for Fig. 3(c). Nevertheless, such decorated boundaries with specified hopping between sites are difficult to achieve in experiment.

Now that we have shown the line states cannot persist in a finite Kagome lattice in general unless the boundaries are carefully engineered, we shall take two alternative ways to demonstrate the NLSs: one is the RBMs spanning the whole boundary of the finite lattice, as proved already in Ref. [38], observing the RBMs is equivalent to observing the NLSs; the other is to use a Kagome lattice in Corbino-geometry for direct observation of the NLSs. Typical experimental and simulation results for the RBMs are presented in Fig. 4, corresponding to the two cases illustrated in Fig. 1(f1) and Fig. 1(f2), replotted in Fig. 4(a) where we shall call the orange line RBM1 and the blue line RBM2. Figure 4(b) is the written Kagome lattice with a lattice spacing of about 34μm. The RBM1 can be considered as a combination of four NLSs by cutting the torus in both toroidal and poloidal directions [38]. To observe this boundary mode, an input probe beam is shaped into a necklace of parallelogram shape, with its phase modulated by an SLM so that adjacent sites have opposite phase. When such a probe beam is launched into the lattice boundary, we see clearly the probe beam preserves after passing through the lattice [Fig. 4(d1)]. For comparison, the necklace deteriorates if the neighboring sites are in-phase [Fig. 4(d2)], as the energy of the boundary mode couples into other sites out of the loop. The RBM2 can be regarded as a linear superposition of RBM1 and a fundamental CLS shown in Fig. 1(a), so there are two defect sites with respect to RBM1 but the loop remains unbroken. Clearly, RBM2 can also be retained after 10-mm transmission through the lattice with the out-of-phase condition [Fig. 4(e1)], but not with the in-phase condition [Fig. 4(e2)].

As necessary, we perform numerical simulations for longer propagation distances to further corroborate the experimental results. We show in Fig. 4 only results from simulation for RBM2 (as results for RBM1 lead to similar conclusion) with parameters from experimental conditions, except for a much longer propagation distance (40 mm). These simulation results show clearly that the RBM remains intact even after a long propagation distance under the out-of-phase condition [Fig. 4(f1)] but becomes strongly distorted under the in-phase condition [Fig. 4(f2)], in excellent agreement with experimental observations. Moreover, as the RBM manifests the realization of the NLS, it is robust against perturbation and even possesses self-healing feature during propagation. In the right panels of Fig. 4, an example is shown for the RBM1, where a phase perturbation (about 25%) is added to one "pearl" of the necklace. After propagating through the lattice, the defect site restores to form a complete boundary mode [Figs. 4(g, h)], which recovers fully for a long-distance transmission (40mm) through the lattice [Fig. 4(i)].

Finally, we design and establish the Corbino-geometry of the photonic Kagome lattice to directly observe the NLS. As illustrated in Fig. 5(a), with this lattice geometry, one can realize the NLS along the toroidal direction, akin to an infinite system in one dimension. For the Corbino-geometry, the distances between $B$ and $C$ sublattices are equivalent over each ring and increase with the ring diameter (here we use $B_1C_1 = C_1B_2 = 32\mu m$); the distances between $A$ and $B$ ($C$) sublattices are also equal but these distances within and outside the ring are not dependent (here we use $A_1C_1 = A_1B_1 = 41\mu m$, and $A_4C_1 = A_4B_2 = 32\mu m$). We generate such a Corbino-shaped Kagome lattice with site-by-site laser writing in the nonlinear crystal, and a typical written lattice is shown in Fig. 5(b). Then, we launch a ring-shaped necklace pattern under out-of-phase condition to excite the NLS depicted in Fig. 5(a). Corresponding experimental results are shown in Fig. 5(c1), where one can clearly see that the necklace beam remains intact after 10mm of propagation, as verified by numerical simulations to even longer propagation [Figs. 5(c2, c3)]. For comparison, if the input necklace beam does not have the required alternating phase, it is strongly distorted during propagation [Figs. 5(d1-d3)] since such an input cannot evolve into the NLS. The results in Fig. 5c represent a direct demonstration of the NLS originally proposed for the infinite system as the FB eigenstate.

In conclusion, we have demonstrated for the first time to our knowledge the RBMs and the NLSs in photonic Kagome lattices, which are unique topological entities available in FB systems. We showed experimentally and theoretically that the line states are not preserved in a finite-sized Kagome lattice since they are not the eigenmodes of the system. Instead, we have taken two alternative approaches to observe an NLS: one is to realize the boundary modes spanning the whole boundary of the finite lattice, which are robust during propagation and exhibit self-healing features, as an indirect realization of the NLS in the finite system; and the other is to achieve direct observation of the NLS in the Corbino-geometry, which hosts the NLS along the toroidal direction. These results are essentially different from those obtained from the Lieb lattices. In fact, since the Lieb lattice represents a bipartite lattice known for the FB ferrimagnetism at half filling while the Kagome lattice serves as an example of a line graph known for the FB ferromagnetism, the fundamental difference between FB ferrimagnetism and ferromagnetism might be visible via the existence or absence of the line states in a finite system. Thus, our work may bring about insights for understanding of these closely related phenomena as well as other intriguing fundamental physics applicable for strongly interacting systems using a convenient photonic platform.

**Acknowledgement:**

This research is supported by the National key R&D Program of China under Grant (No. 2017YFA0303800), National Natural Science Foundation (91750204, 11674180, 11922408 and 11704102), the PCSIRT（IRT0149）and the 111 Project (No. B07013) in China, and the Institute for Basic Science in Korea (IBS-R024-Y1). J.W.R. is supported by IBS-R009-D1, and B.J.Y. by the Institute for Basic Science in Korea (Grant No. 1021 IBS-R009-D1), Basic Science Research Program through the National Research Foundation of Korea (NRF) (Grant Nos. 0426-20170012 and 0426-20180011), and the POSCO Science Fellowship of POSCO TJ Park Foundation (No. 0426- 20180002).

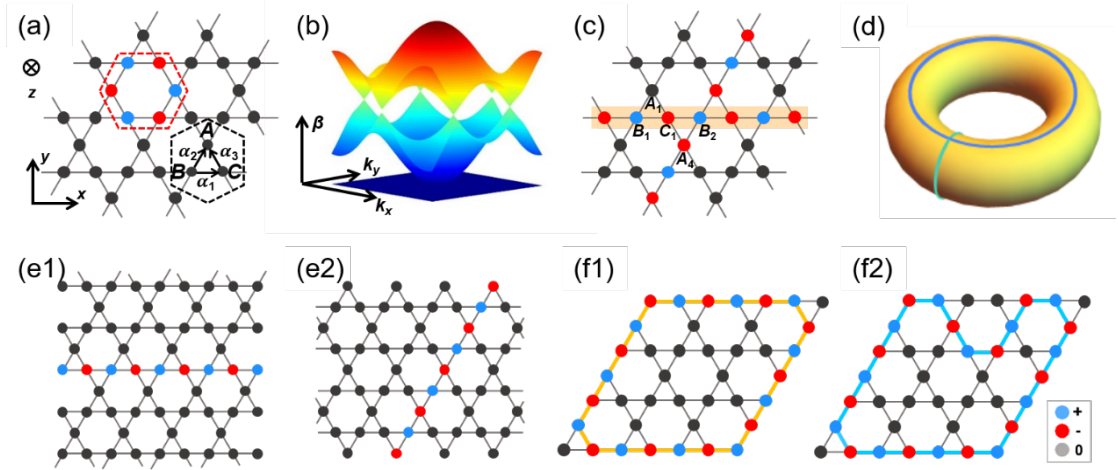

FIG. 1: (a) Schematic of a Kagome lattice, where the black dashed hexagon marks a unit cell with three sublattice sites labeled $A$, $B$, and $C$ and a set of lattice vectors denoted $\alpha_n$ (n=1,2,3) and the red dashed hexagon represents a simplest CLS. (b) Band structure in the tight-binding approximation. (c) Illustration of two NLSs in an infinitely extended Kagome lattice. (d) A torus showing two NLSs mimicking the 2D infinite lattice. (e1, e2) Illustration of line state excitations in finite-sized Kagome lattices with (e1) zigzag-armchair edges at left/right and (e2) armchair edges at top/bottom. (f1, f2) Illustration of two robust boundary modes (orange: RBM1; blue: RBM2) in Kagome lattices with flat cutting edges. In all figures, black sites are of zero-amplitude, while blue and red ones distinguish non-zero sites with opposite phase.

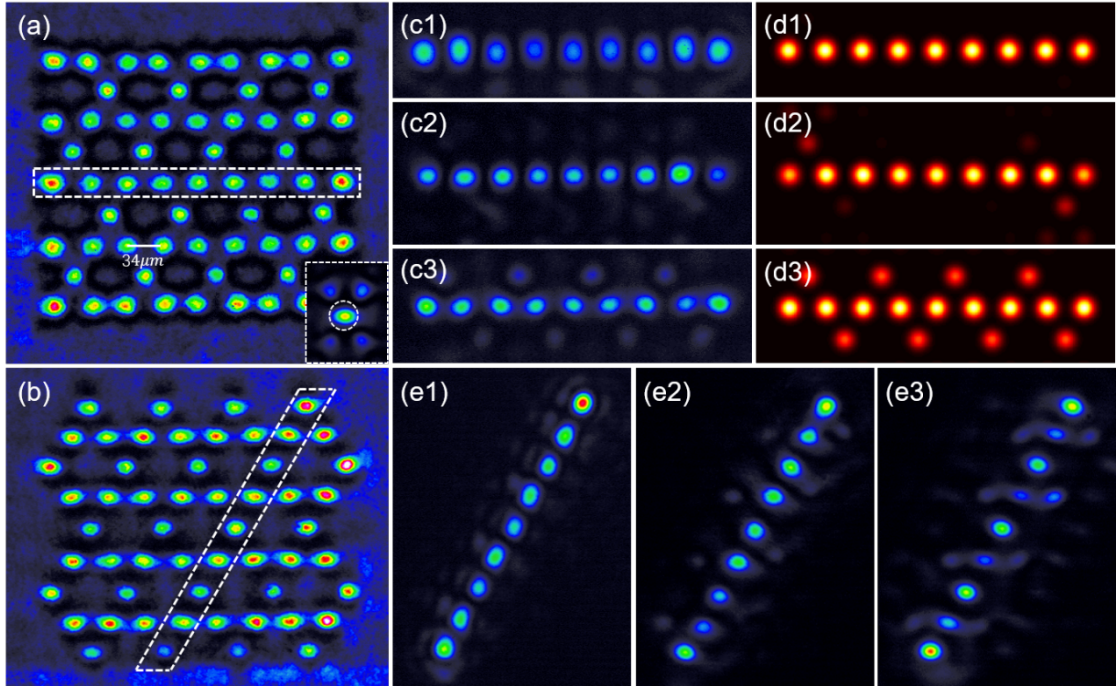

FIG. 2: Experimental realization of finite-sized Kagome lattices with different edges and line state excitation. (a, b) Laser-writing lattices with zigzag-armchair edges located in left and right sides (a) and armchair edges located in top and bottom sides (b), where the white dashed rectangle marks the position of the probing line beam, and the inset in (a) shows discrete diffraction from probing just a single site (marked by the dashed circle) by a Gaussian beam. (c1-c3) Experimental results of the line beam taken at input (c1), and output under out-of-phase (c2) and in-phase (c3) conditions. (d1, d3) Numerical simulations corresponding to (c2, c3) at 10mm propagation, respectively, and (d2) shows diffraction from the ends of the line state after a longer propagation distance of 40mm. (e1-e3) Similar to (c1-c3) but for the excitation corresponding to (b). For lattices in (a) and (b), the nearest neighbor spacing is about 34μm.

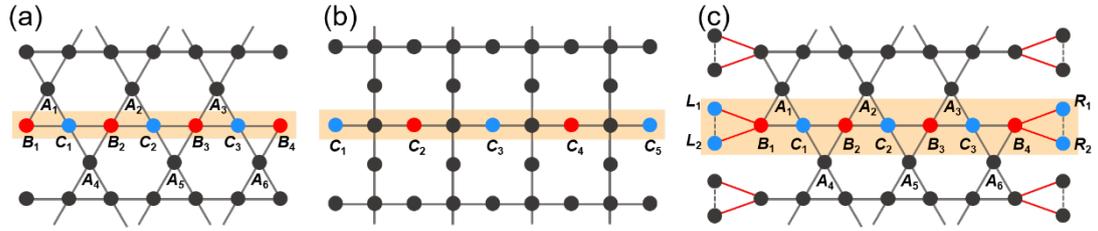

FIG. 3: Comparison of line states in finite-sized Kagome and Lieb lattices. (a) A Kagome lattice with zigzag-armchair edges. (b) A Lieb lattice with bearded edges. (c) A Kagome lattice with decorated boundary formed by attaching two dangling sites at each edge site (denoted by $L$ and $R$), for which the hopping parameters are given by $t$ for black sold lines, $-3t/2$ for black dashed lines, and $t/2$ for red solid lines. Color mapping for all sites is the same as for Fig. 1.

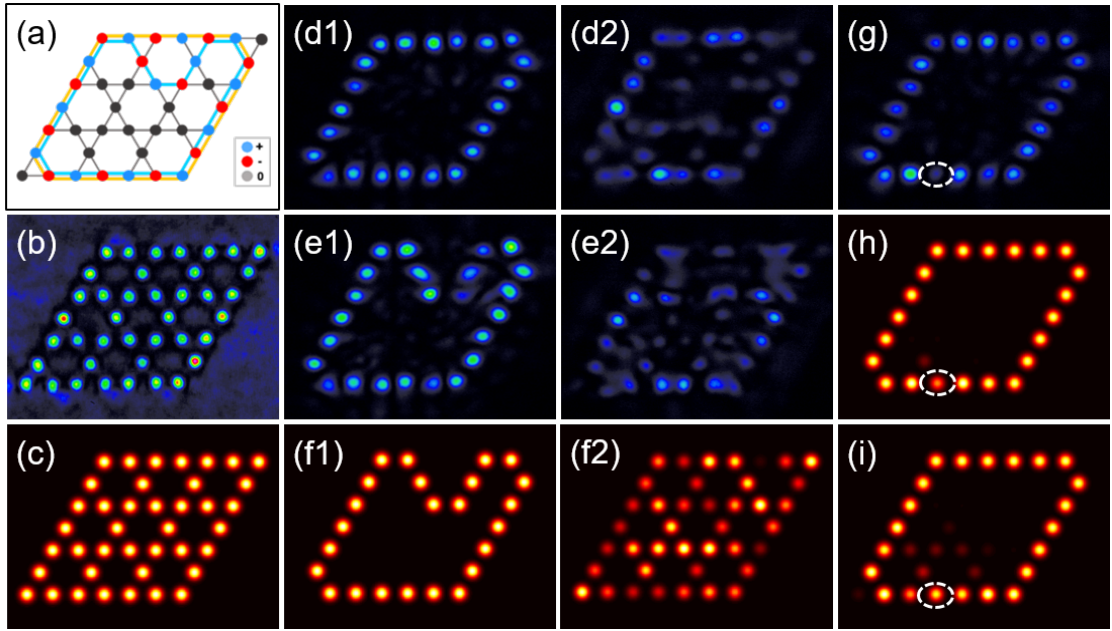

FIG. 4: Realization of RBMs in finite-sized Kagome lattices with flat cutting edges. (a) Illustration of the Kagome lattice and two robust boundary modes (orange: RBM1; blue: RBM2). (b, c) The Kagome lattice obtained from (b) experiment and (c) simulation. (d1, d2) Experimental results of the RBM1 under (d1) out-of-phase and (d2) in-phase conditions. (e1, e2) Corresponding results of the RBM2. (f1, f2) Simulation results corresponding to (e1, e2) but for a much longer propagation distance (40 mm). (g-i) Excitation of the RBM1 with imperfect input phase (25% perturbation is introduced in the defect site circled by dashed line), where (g) is from experiment and (h, i) from simulations at (h) 10mm and (i) 40mm.

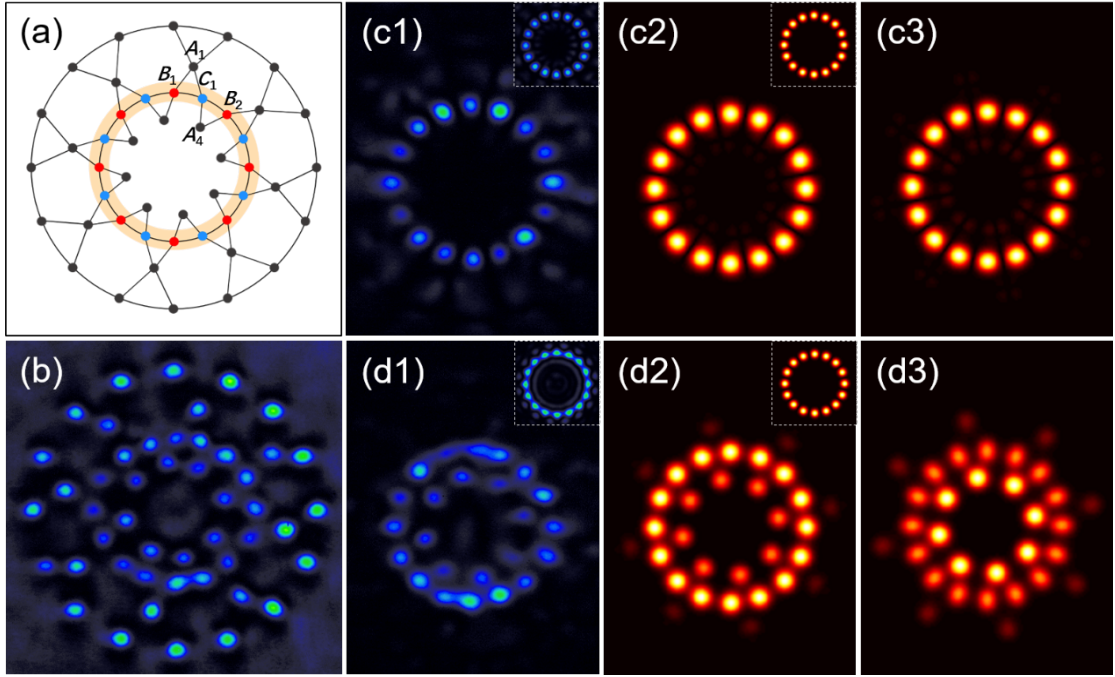

FIG. 5: Realization of non-contractible loop states in finite-sized Kagome lattices in a Corbino-geometry. (a) Schematic diagram of the Corbino-shaped Kagome lattice, where the NLS is illustrated by the orange circle. (b) Experimentally obtained lattice corresponding to (a) by laser-writing. (c1-c3) The NLS observed in (c1) experiment and (c2, c3) simulations at propagation distance of (c2) 10mm and (c3) 40mm under out-of-phase condition. (d1-d3) Corresponding results under in-phase condition. All insets are from input ring necklace of the probe beam.